\documentclass[aps,prb,reprint,superscriptaddress]{revtex4-1}

\usepackage{graphicx,amssymb,bm,amsfonts,amsmath,graphics,color,epstopdf}
\usepackage[bookmarks=false,pdfstartview=FitH,colorlinks=true,citecolor=blue,linkcolor=blue]{hyperref}

\begin{document}

\title{Electronic structure, spin-orbit coupling, and interlayer interaction in bulk MoS$_2$ and WS$_2$}

\author{Drew W. Latzke}
\affiliation{Applied Science \& Technology, University of California, Berkeley, California 94720, USA}
\affiliation{Materials Sciences Division, Lawrence Berkeley National Laboratory, Berkeley, California 94720, USA}

\author{Wentao Zhang}
\affiliation{Materials Sciences Division, Lawrence Berkeley National Laboratory, Berkeley, California 94720, USA}
\affiliation{Department of Physics, University of California, Berkeley, California 94720, USA}

\author{Aslihan Suslu}
\affiliation{School for Engineering of Matter, Transport and Energy, Arizona State University, Tempe, AZ 85287, USA}

\author{Tay-Rong Chang}
\affiliation{Department of Physics, National Tsing Hua University, Hsinchu 30013, Taiwan}

\author{Hsin Lin}
\affiliation{Centre for Advanced 2D Materials and Graphene Research Centre, National University of Singapore, Singapore 117546}
\affiliation{Department of Physics, National University of Singapore, Singapore 117542}

\author{Horng-Tay Jeng}
\affiliation{Department of Physics, National Tsing Hua University, Hsinchu 30013, Taiwan}
\affiliation{Institute of Physics, Academia Sinica, Taipei 11529, Taiwan}

\author{Sefaattin Tongay}
\affiliation{School for Engineering of Matter, Transport and Energy, Arizona State University, Tempe, AZ 85287, USA}

\author{Junqiao Wu}
\affiliation{Materials Sciences Division, Lawrence Berkeley National Laboratory, Berkeley, California 94720, USA}
\affiliation{Department of Materials Science and Engineering, University of California, Berkeley, California 94720, USA}

\author{Arun Bansil}
\affiliation{Department of Physics, Northeastern University, Boston, Massachusetts 02115, USA}

\author{Alessandra Lanzara}
\email[To whom correspondence should be addressed. Email: ]{alanzara@lbl.gov}
\affiliation{Materials Sciences Division, Lawrence Berkeley National Laboratory, Berkeley, California 94720, USA}
\affiliation{Department of Physics, University of California, Berkeley, California 94720, USA}

\date{\today}

\begin{abstract}
We present in-depth measurements of the electronic band structure of the transition metal dichalcogenides (TMDs) MoS$_2$ and WS$_2$ using angle-resolved photoemission spectroscopy, with focus on the energy splittings in their valence bands at the \text{K} point of the Brillouin zone. Experimental results are interpreted in terms of our parallel first-principles computations. We find that interlayer interaction only weakly contributes to the splitting in bulk WS$_2$, resolving previous debates on its relative strength. We additionally find that across a range of TMDs, the band gap generally decreases with increasing magnitude of the valence band splitting, molecular mass, or ratio of the out-of-plane to in-plane lattice constant. Our results provide an important reference for future studies of electronic properties of MoS$_2$ and WS$_2$, and their applications in spintronics and valleytronics devices.
\end{abstract}


\maketitle

\section{\label{sec:Intro}Introduction}
Transition metal dichalcogenides (TMDs) (MX$_2$ where M = Mo or W and X = S, Se, or Te) are layered semiconducting materials that have recently shown promise for application to a wide range of electronic devices, including the growing fields of spintronics and valleytronics\cite{Wang2012c}. In the monolayer limit, TMDs are theorized to have a split valence band that is nearly fully spin polarized\cite{Zhu2011b} near the \text{K} and $\text{K}'$ points of the hexagonal Brillouin zone. The spin polarization is predicted to be reversed between the \text{K} and $\text{K}'$ points, implying existence of an unusual spin-valley coupling in the TMDs. The ability to selectively populate\cite{Cao2012,Xiao2012,Zeng2012} these valleys demonstrates their viability for use in spintronics/valleytronics devices.

Despite the importance of the split valence band that governs the unique spin and valley physics of TMDs, there remain questions regarding the origin of these splittings in bulk TMDs. In particular, this splitting is theorized to be entirely a consequence of spin-orbit coupling in the monolayer limit, and a combination of spin-orbit coupling and interlayer interaction in the bulk limit, but there is disagreement\cite{Klein2001,Molina-Sanchez2013,Alidoust2014,Eknapakul2014} about the relative strength of the two mechanisms in the bulk limit. Current electronic band structure studies of bulk MoS$_2$\cite{Eknapakul2014,Jin2013,Mahatha2012,Mahatha2012a,Mahatha2013,Suzuki2014} and WS$_2$\cite{Klein2001} are limited by energy and momentum resolution and lack focus on the valence band splittings.  Experimentally, the sizes of the splittings have been characterized, but only with limited resolution.

Here we present high-resolution angle-resolved photoemission spectroscopy (ARPES) data of the electronic band structure of bulk MoS$_2$ and WS$_2$, and analyze our measurements via parallel first-principles computations. Our findings resolve previous debates on the role of interlayer interaction on the valence band splitting, revealing that the splitting in the bulk samples is primarily due to spin-orbit coupling and not interlayer interaction. In order to gain insight into the nature of this splitting, we further investigate its origins, its magnitude, and its correlation with the electronic band gap across a range of TMDs. Our comparisons for various TMDs reveal the interplay between the valence band splitting, band gap, molecular mass, and out-of-plane to in-plane lattice constant ratio, suggesting how specific TMDs would be suited for particular applications. 

\section{\label{sec:Met}Methods}
High-resolution ARPES experiments on MoS$_2$ (WS$_2$) were performed at Beamline 10.0.1.1 (4.0.3) of the Advanced Light Source at a temperature of 150~K using 50~eV (52-120~eV) photons. The total energy resolution was 10~meV (20~meV) for experiments on MoS$_2$ (WS$_2$) with an angular resolution ($\Delta\theta$) of $\leq0.2^{\circ}$. The momentum resolution ($\Delta$$k_\parallel$) is related to the angular resolution by $\Delta$$k_\parallel \simeq \sqrt{2mE_{k}/\hbar^2} \cdot \cos \theta \cdot \Delta\theta$ where the kinetic energy of the ejected electrons ($E_{k}$) is dependent on the incident photon energy\cite{Damascelli2003}. The samples were cleaved in situ at 150~K in a vacuum better than $5\times10^{-11}$~Torr.

The electronic structures of MX$_2$ (M = Mo or W and X = S, Se, or Te) were computed using the projector augmented wave method\cite{Bloechl1994,Kresse1999} as implemented in the VASP\cite{Kresse1993,Kresse1996,Kresse1996a} package within the generalized gradient approximation (GGA)\cite{Perdew1996} scheme. A $15 \times 15 \times 3$ Monkhorst-Pack k-point mesh was used in the computations. We used experimental lattice constants and relaxed the atomic positions until the residual forces were less than 0.001~eV/\AA. The spin-orbit coupling effects were included self-consistently. In order to correct the energy band gaps, we also performed calculations with the HSE hybrid functional\cite{Heyd2003,Heyd2006}.

Single crystals were synthesized using the Br$_2$ vapor transport technique in a two-zone furnace system. Mo, W, and S powders (purity 99.9995\%) were mixed in stoichiometric ratios in a sealed quartz tube at $1\times10^{-7}$~Torr pressure and annealed to $880\,^{\circ}\mathrm{C}$ for one week to yield precursor powders. Next, they were loaded into 2 inch diameter tubes with Br$_2$ liquid, pumped down to $1\times10^{-7}$~Torr, and sealed. Typical growth parameters were established as heat up for one week from room temperature to $910\,^{\circ}\mathrm{C}$, the low temperature zone was set to $890\,^{\circ}\mathrm{C}$ in one day, and the system was kept at this temperature for three weeks and controllably cooled down to room temperature in one week. Synthesized crystals displayed strong Raman signals at 384~cm$^{-1}$ and 407~cm$^{-1}$ for MoS$_2$, and 352~cm$^{-1}$ and 417~cm$^{-1}$ for WS$_2$. Lastly, monolayers exfoliated from MoS$_2$ and WS$_2$ crystals displayed sharp bright luminescence peaks at 1.88~eV and 2.04~eV, respectively.

\begin{figure}
\includegraphics[width=\linewidth]{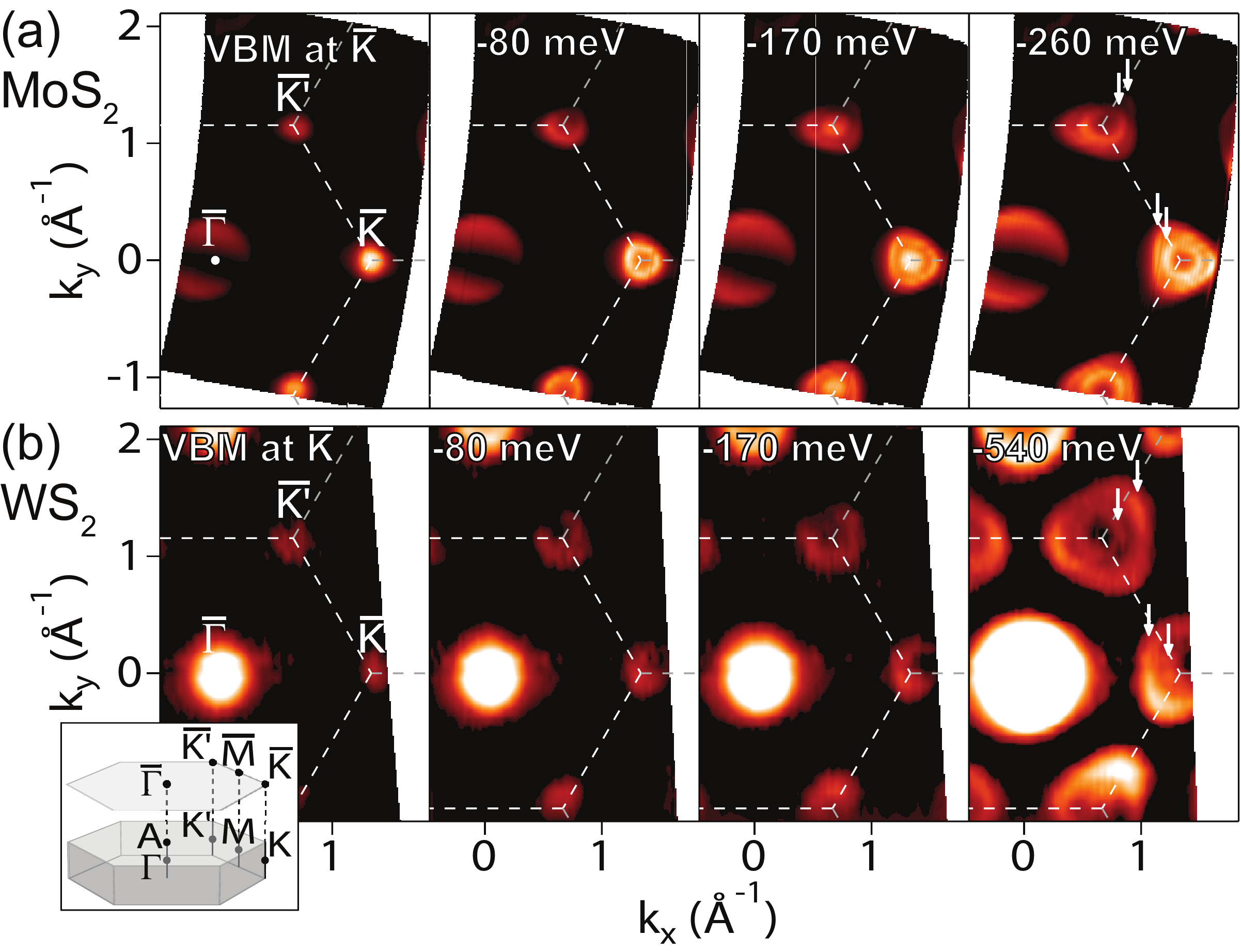}
\caption{\label{fig:ConstE}Constant energy maps of the electronic band structure of bulk \textbf{(a)} MoS$_2$ and \textbf{(b)} WS$_2$ taken at photon energies of 50~eV and 90~eV, respectively. The maps were taken at and below the energy corresponding to their respective valence band maximum (VBM) at $\overline{\text{K}}$. The dashed white line indicates the two-dimensional (reduced in the $k_z$ dimension as shown in the inset) hexagonal Brillouin zone. Evolution of the valence band with increase in binding energy about the $\overline{\Gamma}$, $\overline{\text{K}}$, and $\overline{\text{K}'}$ points can be observed for both samples. An energy splitting at the $\overline{\text{K}}$ and $\overline{\text{K}'}$ points can be observed in the MoS$_2$ maps, most easily at 170~meV or 260~meV below the VBM (white arrows). The splitting for WS$_2$ is not observed 170~meV below the VBM, but can be observed 540~meV below the VBM (white arrows). Trigonal warping of the valence band about the $\overline{\text{K}}$ and $\overline{\text{K}'}$ points can be observed for both samples.}
\end{figure}

\section{\label{sec:Res}Results And Discussion}
\subsection{Characterizing the Origins of the Splitting}
Fig.~\ref{fig:ConstE}(a) and \ref{fig:ConstE}(b) show detailed momentum-resolved constant-energy maps for bulk MoS$_2$ and WS$_2$, respectively. The two-dimensional (reduced in the $k_z$ dimension as shown in the inset) hexagonal Brillouin zone is overlaid as white dashed lines. The observed suppression of intensity in MoS$_2$ along a specific momentum direction near $\overline{\Gamma}$ is due presumably to matrix element effects\cite{Bansil1999,*Sahrakorpi2005,*Bansil2005}. The evolution of the valence band about the $\overline{\Gamma}$, $\overline{\text{K}}$, and $\overline{\text{K}'}$ points can be clearly observed throughout the range of binding energies shown. We offset the two sets of data so that the valence band maximum at $\overline{\text{K}}$ is shown at the same energy and we compare the two band structures as binding energy is increased. The two samples produce similar spectra with the most notable difference being the appearance of the lower split valence band at $\overline{\text{K}}$ at approximately 170~meV below the local valence band maximum (VBM) for MoS$_2$. This appears as a small circular feature centered at $\overline{\text{K}}$ within a larger concentric feature. The analogous splitting for WS$_2$ becomes apparent in the spectra at much higher binding energy, as seen in the map at 540~meV below the VBM. The splitting for both samples is shown by vertical white arrows. Trigonal warping of the valence band in the vicinity of the $\overline{\text{K}}$ and $\overline{\text{K}'}$ points, similar to that seen in other bulk TMDs\cite{Alidoust2014,Suzuki2014}, is observed for both samples and presented here for the first time for bulk WS$_2$. This deviation from a circular feature about $\overline{\text{K}}$ and $\overline{\text{K}'}$ is more pronounced at higher binding energies where the feature becomes more triangular. An analogous trigonal warping has been predicted\cite{Rostami2013,Kormanyos2013} to occur in monolayer MoS$_2$ samples as a consequence of the orbital structure of these bands and anisotropic (order $q^3$) corrections to the energy but has not been observed so far in an ARPES experiment.

\begin{figure}
\includegraphics[width=\linewidth]{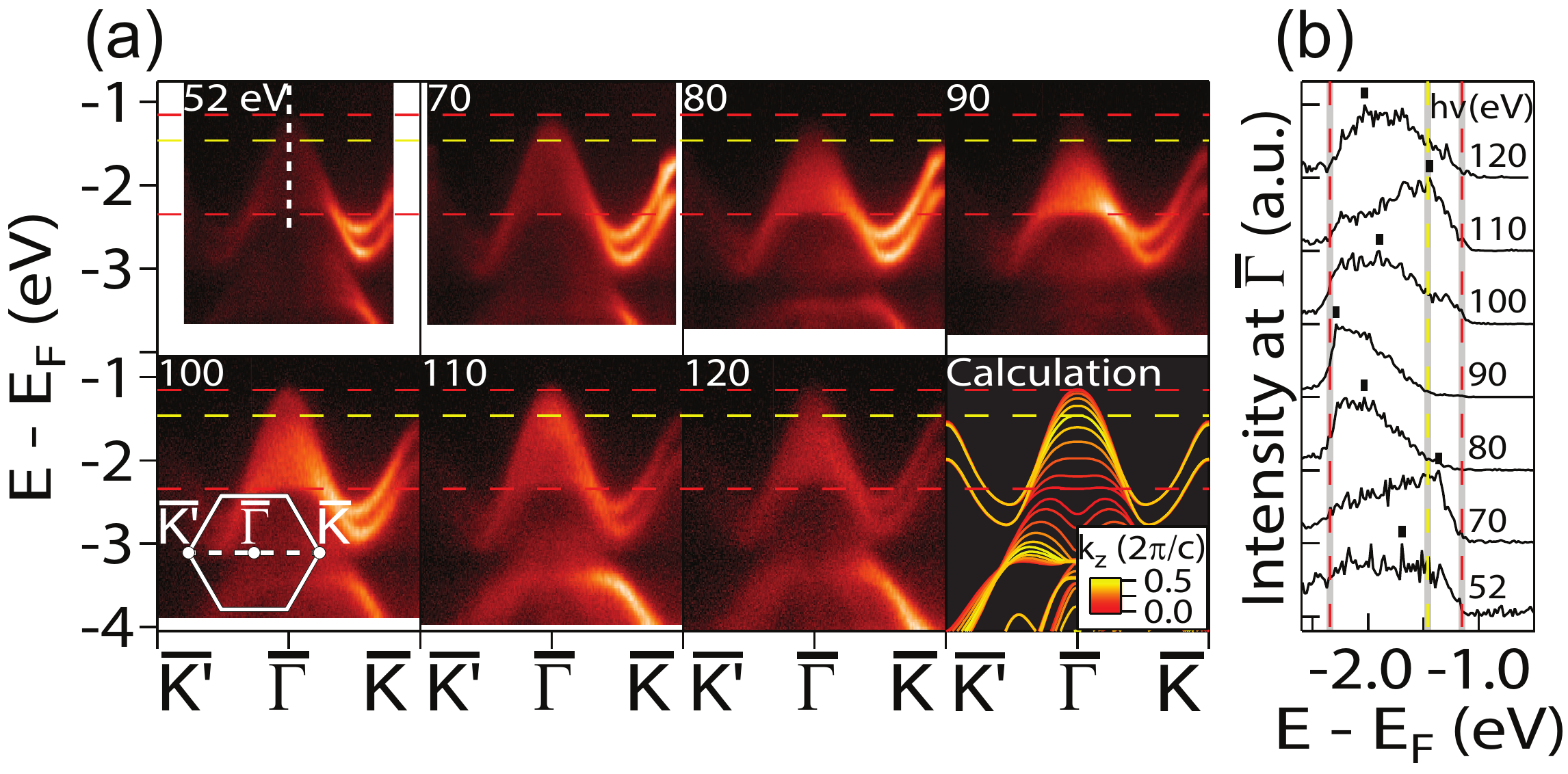}
\caption{\label{fig:WS2PhotDep}\textbf{(a)} Photon energy dependence from 52 to 120~eV and theoretical calculations of the electronic band structure of bulk WS$_2$ along the $\overline{\Gamma}$--$\overline{\text{K}}$ high symmetry direction. The horizontal red (yellow) dashed lines correspond to the calculated location of bulk WS$_2$ bands at the $\Gamma$ point (A point). Observed intensity shifts of the valence band near the $\overline{\Gamma}$ point at different binding energies is evidence of strong $k_z$ dispersion. \textbf{(b)} Energy distribution curves (EDCs) (solid black lines) taken along k$_x$ = k$_y$ = 0~\AA$^{-1}$ of the respective maps in (a) (e.g. along the white dashed line in the 52~eV map). The vertical red (yellow) dashed lines correspond to the calculated location of bulk WS$_2$ bands at the $\Gamma$ point (A point). The peak in intensity (denoted by the tick marks) shifts periodically between higher and lower binding energies as a result of the strong $k_z$ dispersion. Peaks in the 70~eV and 110~eV EDCs as well as the top of the feature in the 90~eV EDC correspond to the theoretical band maximum at A.}
\end{figure}

To obtain information about the $k_z$ dispersion, i.e. the interlayer dispersion, in Figure~\ref{fig:WS2PhotDep}(a) we show detailed photon energy dependent maps compared with $k_z$-resolved theoretical calculations of bulk WS$_2$ along the $\overline{\Gamma}$--$\overline{\text{K}}$ high symmetry direction. Indeed, at normal emission ($k_\parallel = 0$), $k_z$ is related to the kinetic energy of the ejected electrons ($E_k$) by $k_z = \sqrt{2m(E_k+V_{in})/\hbar^2}$ where $V_{in}$ is the inner potential determined from the measured dispersion\cite{Himpsel1983,Hufner2003}. Near $\overline{\Gamma}$, we observe a broad feature likely associated with the convolution of the various bands predicted for different $k_z$ values as it extends between the two horizontal red dashed lines corresponding to the calculated energies of the top two valence bands at $\Gamma$ ($k_z = 0$). The top of this broad feature aligns well with the calculated valence band maximum at $\Gamma$ ($k_z = 0$) for the 52~eV map and as the photon energy is increased further, the top of the feature decreases in energy until it aligns with the calculated valence band maximum at the A point ($k_x = k_y = 0$, $k_z = \frac{\pi}{c}$) indicated by the horizontal yellow dashed line in the 90~eV map. Increasing the photon energy further returns the top of the valence band feature to its maximum value.

The observed $k_z$ dispersion at $\overline{\Gamma}$ is investigated more closely in Fig.~\ref{fig:WS2PhotDep}(b) by comparing energy distribution curves (EDCs) taken at $\overline{\Gamma}$ for each of the incident photon energies (e.g. along the dashed white line in the 52~eV image in (a)). The observation of a large spread in energy of the main peak in the EDC spectra between the calculated energies of the top two valence bands at $\Gamma$ (vertical red dashed lines) is in agreement with the association of the broad feature near $\overline{\Gamma}$ with the convolution of multiple bands predicted for different $k_z$ values. The peak intensity (tick mark) moves up and down in binding energy in a periodic manner confirming the strong $k_z$ dependence of these bands. Overall, our observed $k_z$ dispersions agree well with other predictions for WS$_2$\cite{Klein2001} and our calculations which predict wide variation in binding energy (up to 880~meV) of the bands as $k_z$ changes from 0 to $\frac{\pi}{c}$. Similar $k_z$ dispersions are also found theoretically and experimentally more generally in TMDs\cite{Boeker2001,Fives1992,Mattheiss1973,Coehoorn1987}.

Near $\overline{\text{K}}$, we observe the band splitting at the top of the valence bands for all photon energies, indicating that the splitting occurs for all corresponding $k_z$ values. We measure the magnitude of the splitting to vary between 414~meV and 441~meV in accordance with our corresponding predicted range of 410~meV ($k_z = \frac{\pi}{c}$) to 466~meV ($k_z = 0$). This relatively small change in magnitude of the splitting (27~meV experimentally and 56~meV theoretically) across this range of photon energies indicates weak $k_z$ dispersion near $\overline{\text{K}}$ and $\overline{\text{K}'}$. This is in contrast to the observation of large variations in the spectra of the broad feature at $\overline{\Gamma}$ which indicates strong $k_z$ dispersion of the bands near $\overline{\Gamma}$.

As a result of the layered structure of bulk TMDs, the three-dimensional character of the band structure is directly related to the strength of the interlayer interaction. The strong $k_z$ dispersion at $\overline{\Gamma}$ is a reflection of the three-dimensional character of those bands. However, the relatively weak $k_z$ dispersion at $\overline{\text{K}}$ and $\overline{\text{K}'}$ is a reflection of the two-dimensional character of those bands and shows that interlayer interaction has a weak effect on the valence band splitting. Hence, we conclude that the splitting is dominated by spin-orbit coupling with interlayer interaction playing a minor role.

\begin{figure}
\includegraphics[width=\linewidth]{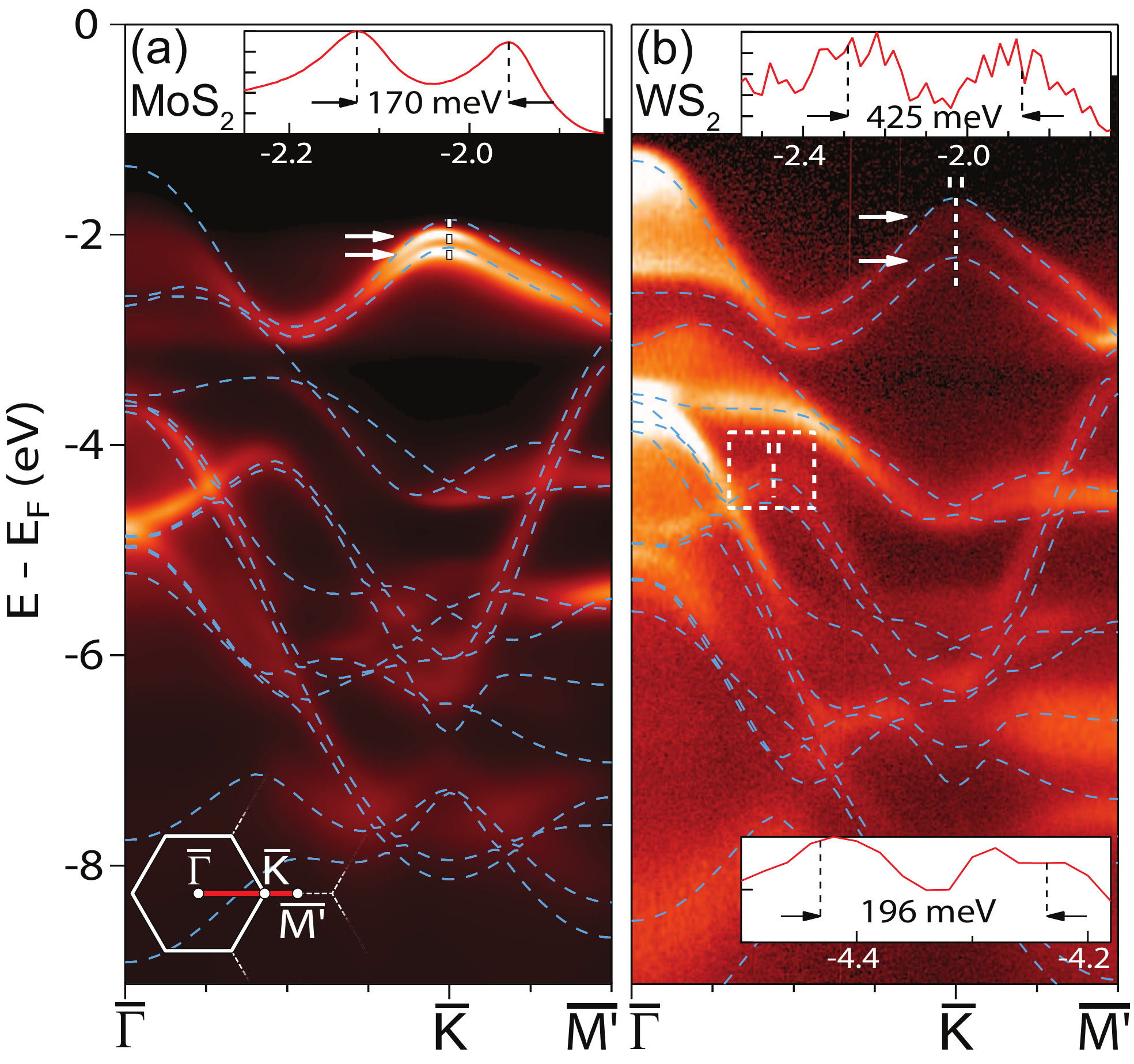}
\caption{\label{fig:Energykplot}High resolution maps of the electronic band structure of bulk \textbf{(a)} MoS$_2$ and \textbf{(b)} WS$_2$ along the $\overline{\Gamma}$--$\overline{\text{K}}$ high symmetry direction with theoretical calculations overlaid as dashed blue lines. Spin-orbit coupling and interlayer interaction induced splitting in the valence band near the $\overline{\text{K}}$ point is fully resolved for both samples as indicated by the white arrows and the respective EDCs (upper insets) taken along the upper white dashed lines. [Here the WS$_2$ EDC was integrated in momentum between the two white tick marks.] The splitting is found to be approximately 170 $\pm$ 2~meV for MoS$_2$ and 425 $\pm$ 18~meV for WS$_2$. An additional splitting in a higher binding energy WS$_2$ valence band is observed within the white dashed rectangle in panel (b). The lower inset presents an EDC taken along the vertical white dashed line corresponding to the energy and momentum location of the maximum splitting of 196 $\pm$ 22~meV. [Here the EDC was integrated in momentum between the two white tick marks and smoothed with a Savitzky-Golay filter.] The calculated band structure is found to be in excellent agreement with the experimental band structure for both samples. The discrepancy in the location of the valence band near the $\overline{\Gamma}$ point for bulk MoS$_2$ can be attributed to its strong $k_z$ dispersion.}
\end{figure}

\subsection{The Effect of Spin-Orbit Coupling on the Valence Band Structure}
Figure~\ref{fig:Energykplot} shows high-resolution energy versus momentum maps of the band structure of bulk MoS$_2$ (panel a) and WS$_2$ (panel b). Band dispersions can be seen as far as 8~eV below the Fermi energy; many of these bands have not been observed before. The experimental band structure shows remarkable agreement with our theoretical predictions (blue dashed lines) throughout this energy and momentum region. The intensity of the valence band maximum near $\overline{\Gamma}$ for MoS$_2$ is reduced due presumably to matrix element effects\cite{Bansil1999,*Sahrakorpi2005,*Bansil2005}. Additionally, intensity of the experimental bands with low binding energy may be shifted near $\overline{\Gamma}$ due to their strong $k_z$ dependence (as shown for WS$_2$ in Fig.~\ref{fig:WS2PhotDep}). The splitting of the valence band at $\overline{\text{K}}$ is fully resolved for both samples (as indicated by the white arrows) and is found to be 170 $\pm$ 2~meV for MoS$_2$ and 425 $\pm$ 18~meV for WS$_2$ (as shown in the upper insets). The value for MoS$_2$ agrees well with our experimental constant energy maps presented in Fig.~\ref{fig:ConstE}(a) as the lower split band for MoS$_2$ appears 170~meV below the higher split band. The value for WS$_2$ is consistent with our experimental constant energy maps in Fig.~\ref{fig:ConstE}(b) as the lower split band is not visible 170~meV below the higher split band, but is clearly visible 540~meV below. These measured magnitudes are smaller by approximately 100~meV than the values of 265~meV for MoS$_2$ and 570~meV for WS$_2$ predicted by HSE-based calculations. However, our experimental values agree very well with GGA-based calculations that predict a splitting of 145-224~meV for MoS$_2$ and 410-466~meV for WS$_2$ across all $k_z$ values. The large difference in magnitude between the splittings for MoS$_2$ and WS$_2$ can be explained by the larger intrinsic spin-orbit coupling of tungsten compared with that of molybdenum. Additionally, our calculations predict a substantial splitting of a valence band of WS$_2$ at high binding energy (see white dashed rectangle). The maximum splitting predicted is in agreement with the experimental value of 196 $\pm$ 22~meV extracted from the EDC spectra (bottom inset) where two clear peaks can be distinguished. Similar to the splitting in the top valence band, we predict this band to be split largely due to spin-orbit coupling and only partially by interlayer interaction.

\begin{figure}
\includegraphics[width=\linewidth]{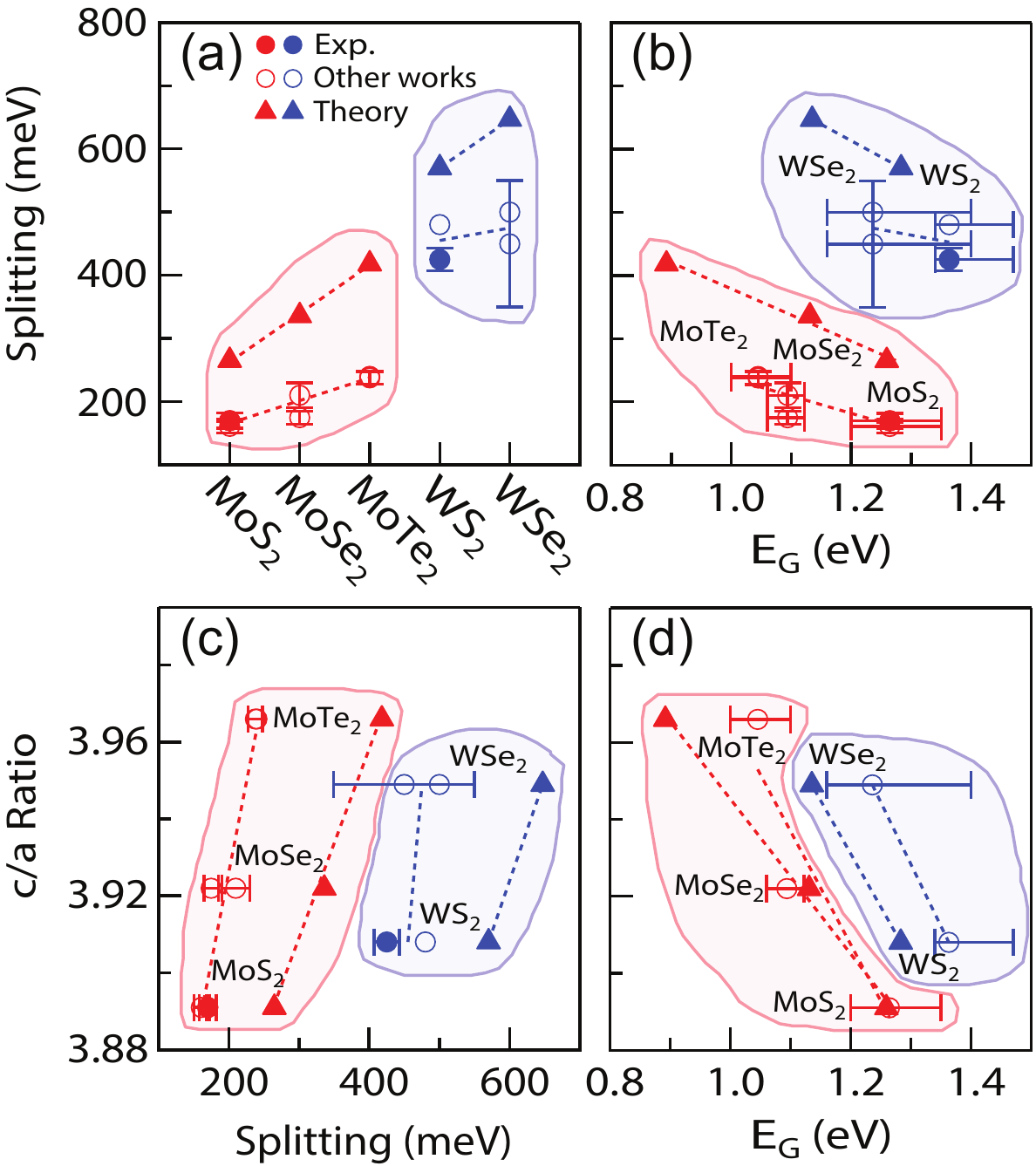}
\caption{\label{fig:ConclusionFig}\textbf{(a)} Magnitude of valence band splitting at the $\overline{\text{K}}$ point for bulk MoS$_2$\cite{Eknapakul2014,Boeker2001}, MoSe$_2$\cite{Coehoorn1987,Boeker2001}, MoTe$_2$\cite{Boeker1999,Boeker2001}, WS$_2$\cite{Klein2001}, and WSe$_2$\cite{Yu1999,Finteis1997}. \textbf{(b)} Magnitude of the splitting vs. band gap (E$_G$) for bulk MoS$_2$\cite{Kam1982,Boeker2001,Aruchamy1992,Gmelin1992}, MoSe$_2$\cite{Kam1982,Boeker2001,Kam1984,Aruchamy1992,Gmelin1992}, MoTe$_2$\cite{Boeker2001,Gmelin1992}, WS$_2$\cite{Braga2012,Kam1982,Aruchamy1992}, and WSe$_2$\cite{Kam1982,Kam1984,Traving1997,Aruchamy1992}. \textbf{(c, d)} Ratio of out-of-plane to in-plane lattice constant ($\frac{c}{a}$)\cite{Boeker2001,Schutte1987} vs. magnitude of valence band splitting (c) and band gap (d). The dashed lines are guides to the eye indicating the different trends for compounds containing molybdenum (red) and those containing tungsten (blue). The trends are further subdivided between experimental (circles) and theoretical (triangles) data. For all plots, filled (open) symbols represent data from this (other) work(s).}
\end{figure}

\subsection{Relationship between Band Gap and Spin-Orbit Splitting}
Figure~\ref{fig:ConclusionFig}(a) shows the magnitude of valence band splitting for bulk MoS$_2$, WS$_2$, and other bulk TMDs. The molybdenum- and tungsten-based TMDs are ordered separately from left to right with increasing molecular mass. For a given transition metal, the magnitude of the splitting increases as molecular mass increases. These trends are shown by the dashed lines and can be explained as a consequence of the larger intrinsic spin-orbit coupling of the successively heavier atoms. The same general trend is obtained from HSE hybrid functional calculations (triangles), although the splitting appears to be approximately 100~meV larger. Figure~\ref{fig:ConclusionFig}(c) shows the ratio of out-of-plane to in-plane lattice constant ($\frac{c}{a}$) versus the magnitude of valence band splitting. Similar to the molecular mass trend, we observe that for a given transition metal, the magnitude of splitting increases with increasing $\frac{c}{a}$ ratio; however, our calculations indicate that although this trend holds across a range of TMDs, it does not hold for individual TMDs. 

In addition to the valence band splitting, the band gap plays a key role when considering TMDs for use in spintronics/valleytronics devices. Figure~\ref{fig:ConclusionFig}(b) shows the magnitude of valence band splitting versus band gap (E$_G$) for bulk MoS$_2$, WS$_2$, and other bulk TMDs. For a given transition metal, the band gap of bulk TMDs decreases with increasing valence band splitting. The same trend is observed as a function of molecular mass in that the band gap decreases with increasing molecular mass. Again, this trend is seen for both experimental and theoretical data even though the magnitude of the splitting for each TMD is calculated to be approximately 100~meV larger than the corresponding experimental values. Figure~\ref{fig:ConclusionFig}(d) shows the ratio of out-of-plane to in-plane lattice constant versus band gap. It too follows the same trend as molecular mass and valence band splitting in that the band gap decreases with increasing $\frac{c}{a}$ ratio. This strong relationship between the band gap, the valence band splitting, and $\frac{c}{a}$ ratio would help guide the search for suitable TMDs targeted at specific applications.

\subsection{Conclusions}
We have presented in-depth high-resolution ARPES studies of the electronic band structure of bulk MoS$_2$ and WS$_2$ together with detailed first-principles computations. We have shown that the valence band splitting in bulk WS$_2$ is primarily due to spin-orbit coupling and not interlayer interaction, resolving previous debates on their relative role. These results strengthen the connection between bulk and monolayer TMDs as the splitting in the monolayer limit is due entirely to spin-orbit coupling. We have shown how the magnitude of this splitting changes for various TMDs and how it correlates with molecular mass, $\frac{c}{a}$ ratio, and band gap. These results provide important new information on how to control these parameters in TMDs and for the development of TMD-based devices in the fields of spintronics and valleytronics.

\begin{acknowledgments}
We thank C. Hwang, C. L. Smallwood, and G. Affeldt for useful discussions. The ARPES work was supported by the sp2 Program at Lawrence Berkeley National Laboratory, funded by the U.S. Department of Energy, Office of Science, Office of Basic Energy Sciences, Materials Sciences and Engineering Division, under Contract No. DE-AC02-05CH11231. The Advanced Light Source is supported by the Director, Office of Science, Office of Basic Energy Sciences, of the U.S. Department of Energy under Contract No. DE-AC02-05CH11231.

The electronic structure calculations at Northeastern University were supported by the US Department of Energy (DOE), Office of Science, Basic Energy Sciences grant number DE-FG02-07ER46352 (core research), and benefited from Northeastern University's Advanced Scientific Computation Center (ASCC), the NERSC supercomputing center through DOE grant number DE-AC02-05CH11231, and support (applications to layered materials) from the DOE EFRC: Center for the Computational Design of Functional Layered Materials (CCDM) under DE-SC0012575. TRC and HTJ are supported by the National Science Council and Academia Sinica, Taiwan. H.L. acknowledges the Singapore National Research Foundation for support under NRF Award No. NRF-NRFF2013-03. We also thank NCHC, CINC-NTU, and NCTS, Taiwan for technical support.
\end{acknowledgments}

\end{document}